\documentclass[preprint2]{aastex}
\usepackage{epstopdf}
\usepackage{amsmath,amssymb}
\usepackage{graphicx,dblfloatfix}
\usepackage[margin=2cm,columnsep=20pt]{geometry}

\begin{document}

\title{Circumstellar Debris Disks: Diagnosing the Unseen Perturber}

\author{Erika R. Nesvold\altaffilmark{1},
Smadar Naoz\altaffilmark{2},
Laura Vican\altaffilmark{2},
Will M. Farr\altaffilmark{3}}
\altaffiltext{1}{Department of Terrestrial Magnetism, Carnegie Institution for Science, 5241 Broad Branch Rd, Washington, DC 20015}
\altaffiltext{2}{Department of Physics and Astronomy, UCLA, 475 Portola Plaza, Los Angeles, CA 90095}
\altaffiltext{3}{School of Physics and Astronomy, University of Birmingham, Birmingham, B15 2TT, UK}

\begin{abstract}

The first indication of the presence of a circumstellar debris disk is usually the detection of excess infrared emission from the population of small dust grains orbiting the star. This dust is short-lived, requiring continual replenishment, and indicating that the disk must be excited by an unseen perturber. Previous theoretical studies have demonstrated that an eccentric planet orbiting interior to the disk will stir the larger bodies in the belt and produce dust via interparticle collisions. However, motivated by recent observations, we explore another possible mechanism for heating a debris disk: a stellar-mass perturber orbiting exterior to and inclined to the disk and exciting the disk particles' eccentricities and inclinations via the Kozai-Lidov mechanism. We explore the consequences of an exterior perturber on the evolution of a debris disk using secular analysis and collisional N-body simulations. We demonstrate that a Kozai-Lidov excited disk can generate a dust disk via collisions and we compare the results of the Kozai-Lidov excited disk with a simulated disk perturbed by an interior eccentric planet. Finally, we propose two observational tests of a dust disk that can distinguish whether the dust was produced by an exterior brown dwarf or stellar companion or an interior eccentric planet. \\

\end{abstract}

\section{Introduction}

Large quantities of dust in debris disks can indicate the presence of an otherwise undetected massive companion in the system \citep[e.g.,][Vican et al. in prep]{Aumann1984, Zuckerman2004, Rhee2007, Lagrange2009, Plavchan2009, Chen2009, Melis2013, Chen2014, Vican2014, Su2014, Meshkat2015}. The dust reported in many of these studies is expected to be short lived, and must either be transient or be continually replenished via collisions excited from interactions with larger bodies. In some cases the system is too young for large bodies to be responsible for the  for this dust. For example, Vican et al. (in prep) showed that self-stirring by particles larger than $\sim1000$ km could be stirring many of the disks, but reported that four of their observed disks are too young to have grown $\sim1000$ km-sized bodies at that distance from their stars. They suggested that in these four disks, a companion is likely gravitationally perturbing the disk and exciting dust-producing collisions.

We consider the effects of a bound perturber on a disk and investigate system architectures that can produce a dust disk. A planet on an eccentric orbit interior to a disk can excite the eccentricities of the disk particles via secular perturbations. For example, \citet{Mustill2009} demonstrated that dynamical excitation via secular perturbations from a giant planet a few AU from its star can cause destructive collisions in a disk out to hundreds of AU. In fact, many directly imaged exoplanets are found in systems with observed debris disks \citep[e.g.,][]{Marois2008,Kalas2008,Lagrange2010}. Thus, the detection of a circumstellar dust population can prompt a direct imaging search for companions orbiting interior to the dust disk.

While the secular effects of an eccentric planet on its disk are well-studied \citep[e.g., see][]{Wyatt1999a,Wyatt2005a, Chiang2009, Boley2012, Pearce2014, Nesvold2015a}, here we focus on an alternative, less-studied scenario: a bound perturber orbiting exterior to the disk with a high inclination ($\gtrsim 40^{\circ}$). Such a perturber will excite large eccentricities in the disk via the eccentric Kozai-Lidov mechanism \citep[e.g., see][]{Kozai1962, Lidov1962, Innanen1997, Naoz2016}, increasing the rate of dust-producing collisions. \citet{Martin2014} and \citet{Fu2015} used hydrodynamical simulations to show that the Kozai-Lidov mechanism can also operate in a disk with gas pressure and viscosity, in the absence of disk self-gravity, for a wide range of system parameters. \citet{Batygin2012}, and \citet{Lai2014} demonstrated that in high-mass disk, the self-gravity of the disk will suppress Kozai-Lidov perturbations, and the disk will precess as a rigid flat body. \citet{Fu2015a} determined that the disk mass required for suppressing Kozai-Lidov perturbations is a few percent the mass of the star. In this work, we consider the effects of Kozai-Lidov perturbations on a gas-free debris disk. 

Systems with perturbers orbiting exterior to a debris disk have already been observed. For example, \citet{Mawet2015} recently reported the discovery of a distant $15-80$~M$_{\rm Jup}$ companion to HR 3549, an A0V that harbors a debris disk. Another example of a disk with a wide companion is the HD 106906 system, which harbors both a debris disk and an extremely distant planet \citep{Bailey2014, Kalas2015, Lagrange2016}. \citet{Jilkova2015} modeled the evolution of the HD 106906 disk under the influence of the distant companion and found that some planet-disk configurations produced a ``wobble'' in the disk, which they attributed to Kozai-Lidov-like oscillations. Here we extend beyond the N-body models of \citet{Jilkova2015} to include disk self-gravity and collisions which, as we will show, play a major role in determining the final configuration of a disk under the influence of a distant, exterior companion.

Here, we explore the formation of a dust disk via the Kozai-Lidov mechanism by simulating a debris ring perturbed by a distant brown dwarf. We also propose observational methods for distinguishing between a dust disk that has been excited by an inclined perturber orbiting exterior to the disk and a dust disk excited by an eccentric planet orbiting interior to the disk. We begin by analyzing the scenario with a disk perturbed by a Kozai-Lidov companion. In Section \ref{sec:analytic}, we derive and compare the collisional and Kozai-Lidov perturbations timescales for such a disk. In Section \ref{sec:methods}, we describe the two numerical methods we used to simulate the evolution of a disk with a perturber. In Section \ref{sec:results}, we present the results of our simulations and compare the results for a disk with an exterior brown dwarf and an interior eccentric planet. In Section \ref{sec:discussion}, we discuss the implications for observers of dust disks, and in Section \ref{sec:summary}, we summarize our results and conclusions.

\section{Analytical Treatment} 
\label{sec:analytic}

Consider a disk of particles (each with radius $D$) on initially circular orbits around a star of mass $M_*$. If we introduce a companion of mass $M_{\rm p}$, eccentricity $e_{\rm p}$, orbital period $P_{\rm p}$, and inclination $i_{\rm p}$ relative to the disk, it will excite the eccentricities and inclinations of the particles via the eccentric Kozai-Lidov mechanism if the perturber's inclination lies in the range  $-\sqrt{3/5} \leq \cos i_{\rm p} \leq \sqrt{3/5}$, or $39.2^{\circ} \lesssim i_{\rm p} \lesssim 140.77^{\circ}$ \citep[see, e.g.][for a detailed description of the Kozai-Lidov mechanism]{Naoz2016}. The timescale of the eccentricity excitation is
\begin{equation} \label{eq:tKozai} t_{\rm Kozai} \sim \frac{M_{\rm tot}}{M_{\rm p}} \frac{P_{\rm p}^2}{P} (1-e_{\rm p}^2)^{3/2}, \end{equation}
where $P$ is the orbital period of the particle and $M_{\rm tot} = M_* + M_{\rm p} + M_{\rm disk}$ is the total mass of the system, including the star, the perturber, and the disk \citep{Antognini2015, Naoz2016}. 
Because $t_{\rm Kozai}$ depends on $P$, the timescale of the eccentricity excitations will differ for different regions of the disk. The maximum eccentricity achieved by the particles in the disk is given by
\begin{equation} \label{eq:emax} e_{\rm max} \approx \sqrt{1-(5/3) \cos^2 i_{\rm p}} \end{equation}
in the quadrupole approximation, the lowest order of the hierarchal three-body approximation \citep{Naoz2013}.

Now suppose that the particles in the disk interact via collisions. Consider an annulus in the disk of width $\Delta a$ at semimajor axis $a$. The collision rate, $\Gamma$, of the particles in the annulus can be estimated by 
\begin{equation} \label{eq:collrate} \Gamma \approx n \sigma v_{\rm rel}, \end{equation}
where $n$ is the number density of the annulus, $\sigma$ is the geometrical cross section for interaction, and $v_{\rm rel}$ is the relative velocity between two particles. The geometrical cross section is given by $\sigma=4\pi D^2$. The number density of particles can be estimated at semimajor axis $a$ as $n\approx N_{\rm ann}/(2\pi a \Delta a h)$, where $N_{\rm ann}=M_{\rm ann}/m$ is the total number of particles of mass $m$ in the annulus, $M_{\rm ann}$ is the total mass in the annulus, and $h$ is the scale height of the disk. If we approximate the particles as solid spheres of density $\rho$, the mass of a particle can be written as $m=4\pi \rho D^3/3$.

We can approximate the relative velocity of two particles on neighboring orbits within the annulus with semimajor axis $a_1\approx a_2 \approx a$ and eccentricities $e_1$ and $e_2$ as
\begin{equation} v_{\rm rel}\approx \sqrt{\frac{G M_\star}{a}}\left(\sqrt{\frac{1-e_1}{1+e_1}}-\sqrt{\frac{1-e_2}{1+e_2}} \right), \end{equation}
where $G$ is the gravitational constant. Since $a_1\approx a_2$, the timescale for eccentricity excitation via the Kozai-Lidov mechanism is approximately equal for the two particles, so their eccentricities should be very similar and we can approximate $e_1 \approx e_2 + \delta e$ for some small $\delta e$. The incremental relative velocity due to this small difference in eccentricities is then
\begin{equation} \delta v_{\rm rel} \approx \sqrt{\frac{GM_\star}{a}} \sqrt{\frac{1-e}{1+e}} \frac{\delta e}{(1+e)(1-e)}. \end{equation}
Both particles will reach the same maximum eccentricity $e_{\rm max}$ under the influence of Kozai-Lidov perturbations from the companion, so we can integrate over eccentricity to $e_{\rm max}$ to find the relative velocity between particles on adjacent orbits:
\begin{eqnarray} \label{eq:eccintegral}
 v_{\rm rel}&\approx&\sqrt{\frac{G M_\star}{a}}\int_0^{e_{\rm max}}\sqrt{\frac{1-e}{1+e}}\frac{\delta e}{(1+e)(1-e)} \nonumber \\
 &=&\sqrt{\frac{G M_\star}{a}}\left(1-\sqrt{\frac{1-e_{\rm max}}{1+e_{\rm max}}} \right) \ ,
\end{eqnarray}
Plugging Equation (\ref{eq:eccintegral}) into (\ref{eq:collrate}) along with our estimates for $n$ and $\sigma$, we find that the collision rate in the disk at semimajor axis $a$ is
\begin{equation} \Gamma\sim \frac{3M_{\rm disk}}{2\pi\rho} \frac{1}{D}\frac{1}{\Delta a h}\sqrt{\frac{G M_\star}{a^3}}\left(1-\sqrt{\frac{1-e_{\rm max}}{1+e_{\rm max}}} \right). \end{equation}
The collisional timescale for the disk is then $t_{\rm coll} \sim \Gamma^{-1}$. We summarize the disk and perturber parameters used in our simulations in Table \ref{tab:initial}.

\begin{deluxetable}{lccc}
\tablewidth{0pt}
\tablecaption{Initial conditions of the disk and perturber for the simulations. \label{tab:initial}}
\tablehead{\colhead{Parameter	} & \colhead{Initial Disk Values} & \colhead{Exterior Perturber} & \colhead{Interior Perturber}}
\startdata
	Semi-Major Axis (AU) 		& ($220-250$)\tablenotemark{a}, ($150-250$)\tablenotemark{b} 		& 800 	& 123 \\
	Eccentricity 				& $0.0-0.01$ 		& 0.2  	& 0.14\\
	Inclination (deg) 			& $0.0-0.29$		& 45 		& 0 \\
	Mass ($\hbox{M}_{\odot}$)	& ($3\times10^{-5}$)\tablenotemark{a} 	& 0.04 	& 0.001 \\
	Optical depth				& ($5\times10^{-4}$)\tablenotemark{b} 	& --		& -- \\
	Scale Height (AU)	 		& 10		 		& -- 		& -- \\
	Density (g~cm$^{-3}$) 		& 1.0	 			& -- 		& -- \\
\enddata

\tablenotetext{a}{For ring simulation only}
\tablenotetext{b}{For SMACK simulation only}

\end{deluxetable}

\begin{figure}[ht]
	\centering
	\includegraphics[width=\columnwidth]{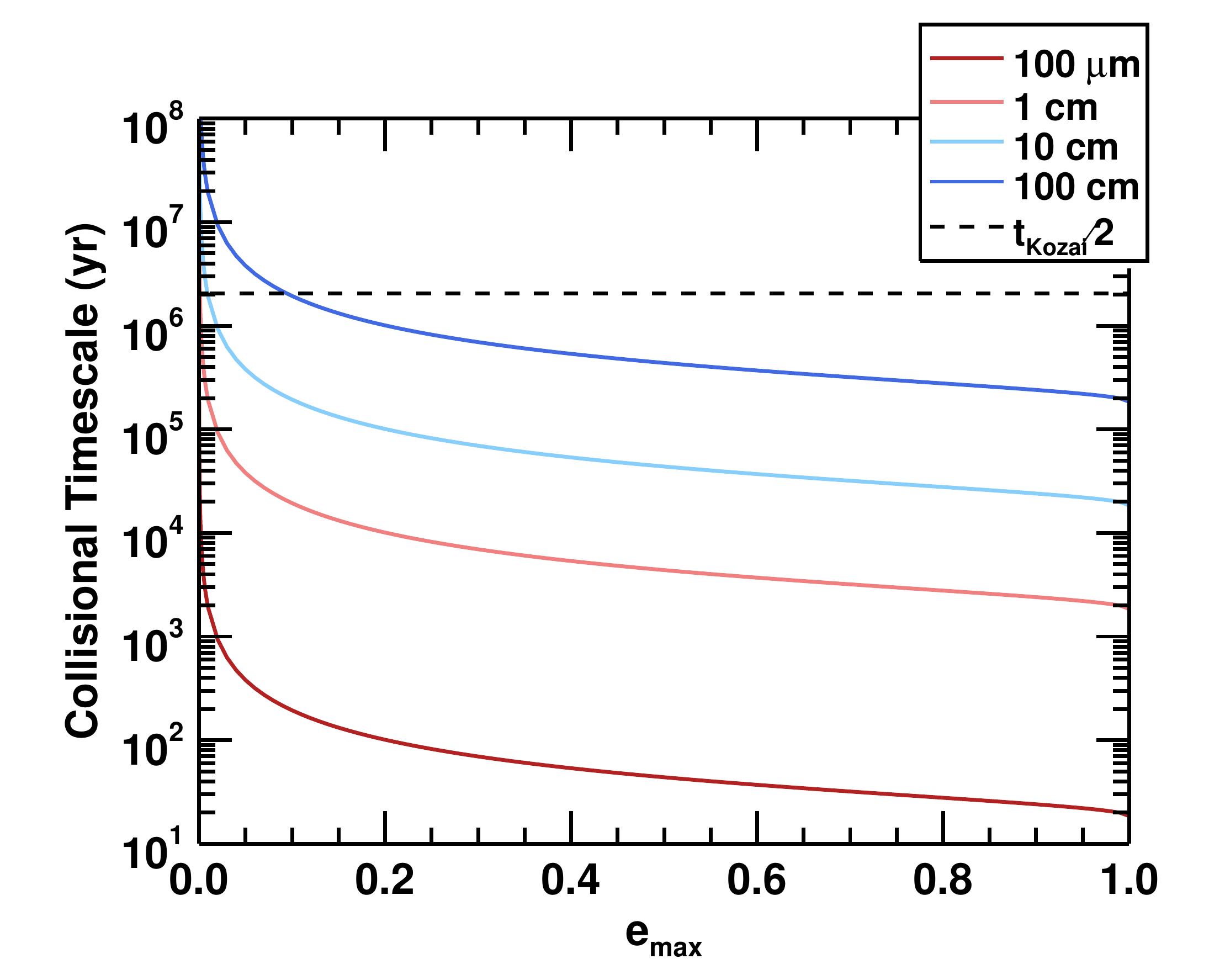}
	\caption{The collisional timescales of different grains in our model disk vs. the maximum eccentricity excited by the perturber. We also plot the Kozai-Lidov timescale divided by half, to indicate the time at which the disk will have reached its first eccentricity peak. For grain sizes $\sim10-100$ cm, collisions and Kozai-Lidov perturbations will both have a significant affect on the dynamics of the disk.}
	\label{fig:analytical}
\end{figure}

Fragmenting collisions between particles in the disk are inelastic and will tend to damp their eccentricities. If the Kozai-Lidov timescale is much longer than the collisional timescale ($t_{\rm Kozai} >> t_{\rm coll}$), collisions will dominate the dynamical evolution of the disk. If $t_{\rm Kozai} << t_{\rm coll}$, the Kozai-Lidov perturbations of the disk will result in extremely large eccentricities that will not only lead to a collisional cascade, but may also disperse the disk. In this work, we are concerned with the regime in which the Kozai-Lidov timescale is similar to the collisional timescale ($t_{\rm Kozai} \sim t_{\rm coll}$), and the eccentricity excitation saturates. In Figure \ref{fig:analytical} we consider the effects of the ``Exterior Pertuber'' described in Table \ref{tab:initial} on a disk with M$_{\rm disk}=3\times10^{-6}$~M$_{\odot}$, $a=200$~AU, $\Delta a=50$~AU, and $h=0.05a$. We plot $t_{\rm coll}$ for different disk particle sizes versus the maximum eccentricity excited by the perturber. We also plot $t_{\rm Kozai}/2$, the time it takes for the disk to reach its first eccentricity peak, which is independent of $e_{\rm max}$. Figure \ref{fig:analytical} indicates that for particles of size $\sim10-100$ cm, $t_{\rm Kozai}\sim t_{\rm coll}$, and the effects of both inelastic collisions and eccentricity excitations on the dynamics of the disk must be considered.

\section{Numerical methods}
\label{sec:methods}

We studied the model disk and perturber described in Section \ref{sec:analytic} with two independent numerical methods: an approximation of the disk as a series of concentric rings, interacting gravitationally, and an N-body simulation that includes fragmenting, inelastic collisions between particles. These two types of models simulate two different types of particle interactions (self gravity vs. inelastic collisions) that may counteract the Kozai-Lidov mechanism.

\subsection{Gravitationally Interacting Rings via Gauss's Averaging Method}
\label{sec:rings}

To study the effects of self-gravity within the disk we employed Gauss's averaging method, a phase-averaged calculation for which the gravitational interactions between non-resonant orbits are equivalent in treating the orbits as massive wires interacting with each other,  where the line-density is inversely proportional to orbital velocity. We calculate the forces that different wires exert on each other. A consequence of the secular approximation is that the semi-major axes of the wires are constants of motion \citep[e.g.,][]{Murray1999}.

Recently, \citet{Touma2009} extended the averaging method to softened gravitational interactions, which is equivalent to dispersing the mass over a Plummer potential. The strength of this approach and thus the code is that it is not limited to low inclinations and eccentricities or hierarchical configurations. As long as the system is not in resonance and the wires do not intersect each other, this calculation is robust. Our code calculates the gravitational interactions between $n>1$ massive wires. This code has been tested compared to N-body integrations.

Using this method, we modeled the disk as 30 rings, each with mass $1/3$~M$_{\rm Earth}$, distributed between 220 and 250 AU. We included a 40~M$_{\rm Jup}$ brown dwarf orbiting exterior to the disk at 800~AU and evolved the system for 14~Myr. The initial parameters of the disk and the exterior brown dwarf perturber are listed in Table \ref{tab:initial}.

\subsection{Collisional Modeling with SMACK}
\label{sec:smack}

To study the effects of particle-particle collisions on the evolution of a disk under the influence of a massive perturber, we used the Superparticle-Method Algorithm for Collisions in Kuiper belts (SMACK), which uses N-body integration\footnote{SMACK uses the REBOUND package for N-body integration \citep{Rein2012}.} to track the orbits of swarms of bodies with a range of sizes, known as ``superparticles'', under the influence of gravitational perturbations from a star and any companions \citep{Nesvold2013}. Unlike the rings in the simulation discussed in Section \ref{sec:rings}, the superparticles do not interact gravitationally with each other. However, when two superparticles overlap, SMACK calculates the probability of fragmenting collisions between bodies in the two superparticles and adjusts the size distributions and trajectories of the superparticles to simulate the loss of kinetic energy and the redistribution of fragments in the size distributions. In this way, SMACK simultaneously evolves the dynamics and size distribution of the 3D disk. We selected the updated collision resolution algorithms described in \citet{Nesvold2015}.

SMACK allows the user to set the initial vertical optical depth of the disk rather than its total mass. We set the initial optical depth to $5\times10^{-4}$, and used 10,000 superparticles for a 30 Myr simulation. We simulated the evolution of the disk under the influence of the same $40$~M$_{\rm Jup}$ perturber at 800 AU that we modeled with the ring simulation (Section \ref{sec:rings}). 

To explore the differences between a dust disk excited by Kozai-Lidov perturbations from an exterior perturber and one excited by an interior, eccentric perturber, we also ran a SMACK simulation of the same disk for 30 Myr, but with a $10$~M$_{\rm Jup}$ planet orbiting coplanar with the disk at 123 AU with eccentricity 0.14. The semi-major axis and eccentricity of the planet were chosen to force an eccentricity of $\sim0.1$ on the disk to simulate a moderately eccentric debris disk in the regime of Fomalhaut \citep{Kalas2013} or $\epsilon$ Eridani \citep{Greaves2014}. The initial conditions of the disk and the perturber (for both the exterior and interior perturber simulations) are summarized in Table \ref{tab:initial}.

\section{Results}
\label{sec:results}
	
\subsection{Perturbation of the Disk Structure}

\begin{figure}[t]
	\centering
	\includegraphics[width=\columnwidth]{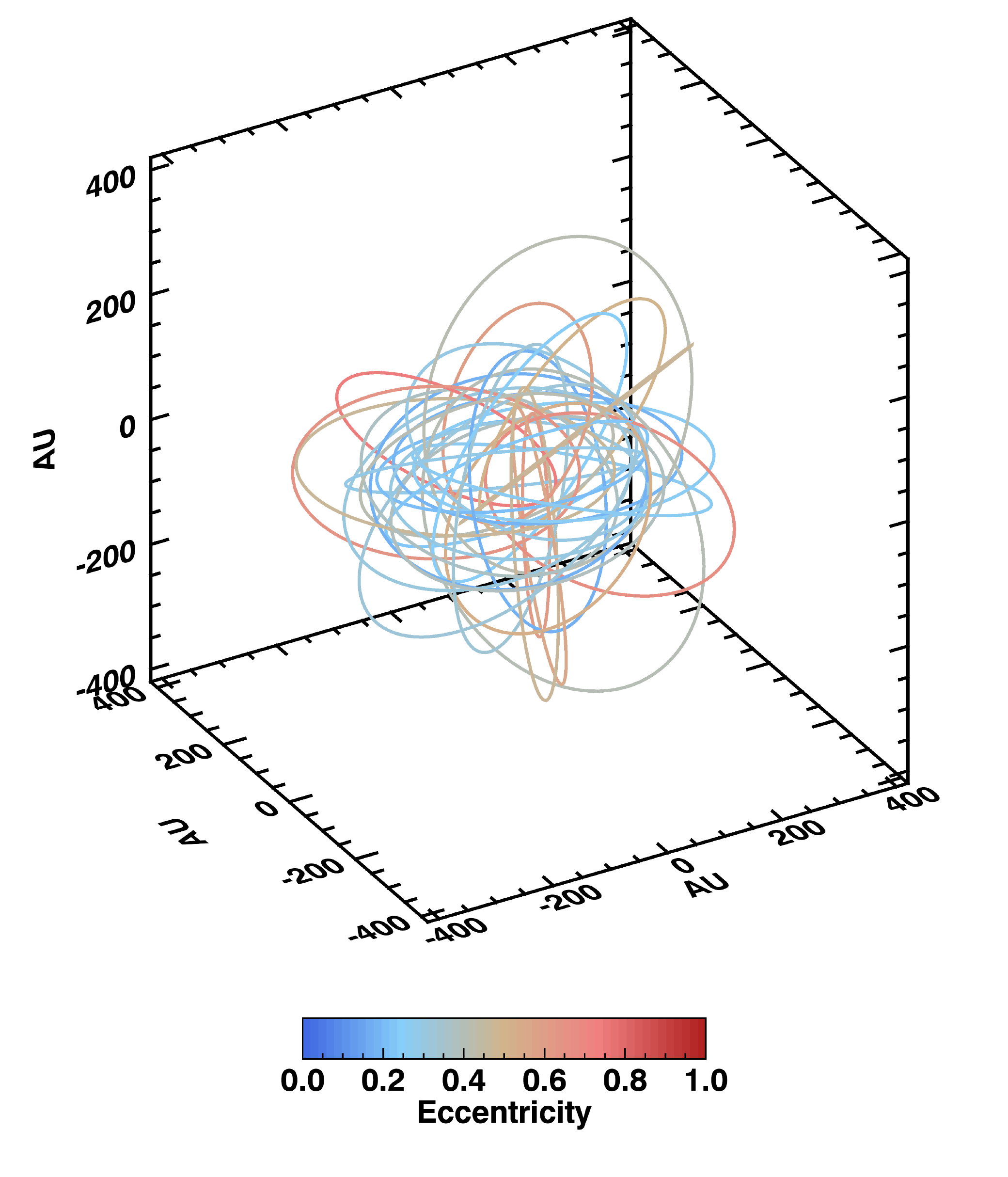}
	\caption{3D structure of the ring-simulated disk under the influence of an inclined exterior perturber after 14 Myr. Colors indicate the eccentricity of the rings.}
	\label{fig:rings3d}
\end{figure} 

Gravitational perturbations of the disk particles' eccentricities and inclinations via the Kozai-Lidov mechanism will alter the disk's structure. Figure \ref{fig:rings3d} shows the 3D orbit configuration of the rings perturbed by the exterior perturber after 14 Myr, with colors indicating the eccentricity of the rings. The exterior perturber has significantly changed the 3D structure of the disk as well as exciting the orbital eccentricities. 

\begin{figure*}[t]
	\centering
	\includegraphics[width=\linewidth]{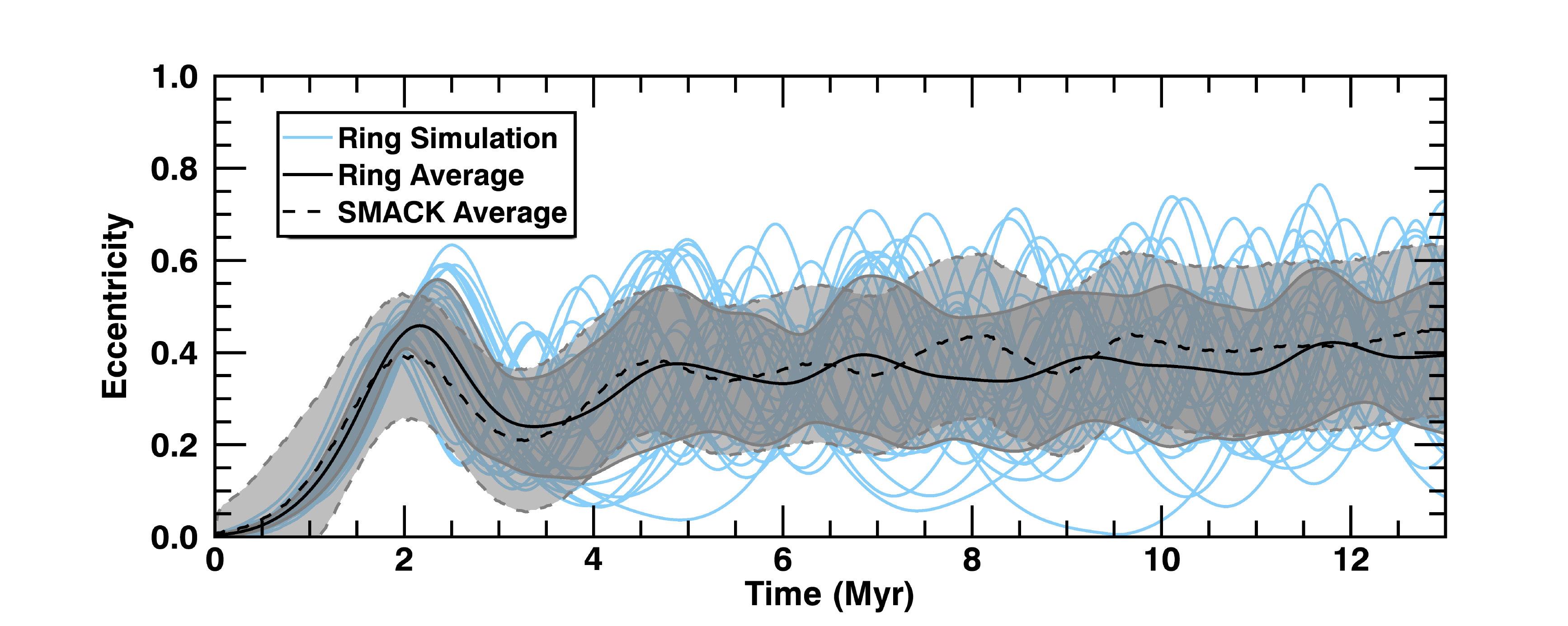}
	\caption{Mean eccentricity of the disk over time in the presence of the 40 $M_{Jup}$ perturber, for both the ring simulation (solid black line) and the SMACK simulation (dashed black line). The eccentricity of each ring in the ring simulation is shown in blue. The grey shaded regions outlined with solid and dashed grey lines indicate the 1-$\sigma$ boundary for the mean eccentricity of the ring and SMACK simulations, respectively. The mean eccentricity peaks after $P_{Kozai}/2\sim 2$~Myr in both simulations, as expected, although the period of the eccentricity oscillation differs slightly between the ring simulation and the SMACK simulation.}
	\label{fig:eccvstime}
\end{figure*}

To compare the eccentricity evolution of the ring model and the SMACK-simulated disk, in Figure \ref{fig:eccvstime} we plot the time evolution of the mean eccentricity of the superparticles in the SMACK simulation with an exterior perturber and the eccentricity of each ring in the ring simulation. 

In both simulations, the eccentricity peaks after $P_{\rm Kozai}/2\sim 2$~Myr and continues to oscillate. While the mutual gravitational interactions between the rings pump their peak eccentricities up to $> 0.7$, the eccentricities of the various rings begin to deviate significantly after the first peak, as the timescale of the Kozai-Lidov excitations depends upon the semi-major axis of the ring. This effect results in an average ring eccentricity that oscillates around a value of $\sim0.4$, close to the maximum eccentricity predicted by Equation \ref{eq:emax}, $e_{max}\sim0.41$. Increasing the width of the modeled disk (i.e., increasing the number of rings) would eventually result in a constant average eccentricity representing the mean of the maximum and minimum eccentricities in the rings. A similar effect occurs in the SMACK-simulated disk: the eccentricities of superparticles at different semi-major axes evolve at different timescales, resulting in an average eccentricity similar to that of the ring model, although the period of the oscillation of the mean eccentricity differs slightly between the ring and SMACK simulations. 

Note again that the ring code and SMACK simulate two different types of interactions between particles in the disk: the ring code simulates the gravitational interactions between particles in a massive, self-gravitating disk with no collisions, and SMACK simulates the damping effect of inelastic collisions between particles in a less massive disk with negligible self-gravity. However, Figure \ref{fig:eccvstime} indicates that in both simulations, the growth in eccentricity predicted by the eccentric Kozai-Lidov mechanism \citep{Naoz2016} is suppressed, by collisional damping in the SMACK simulation and by gravitational interactions between the rings in the ring code. 

\begin{figure*}[t]
	\centering
	\includegraphics[width=\linewidth]{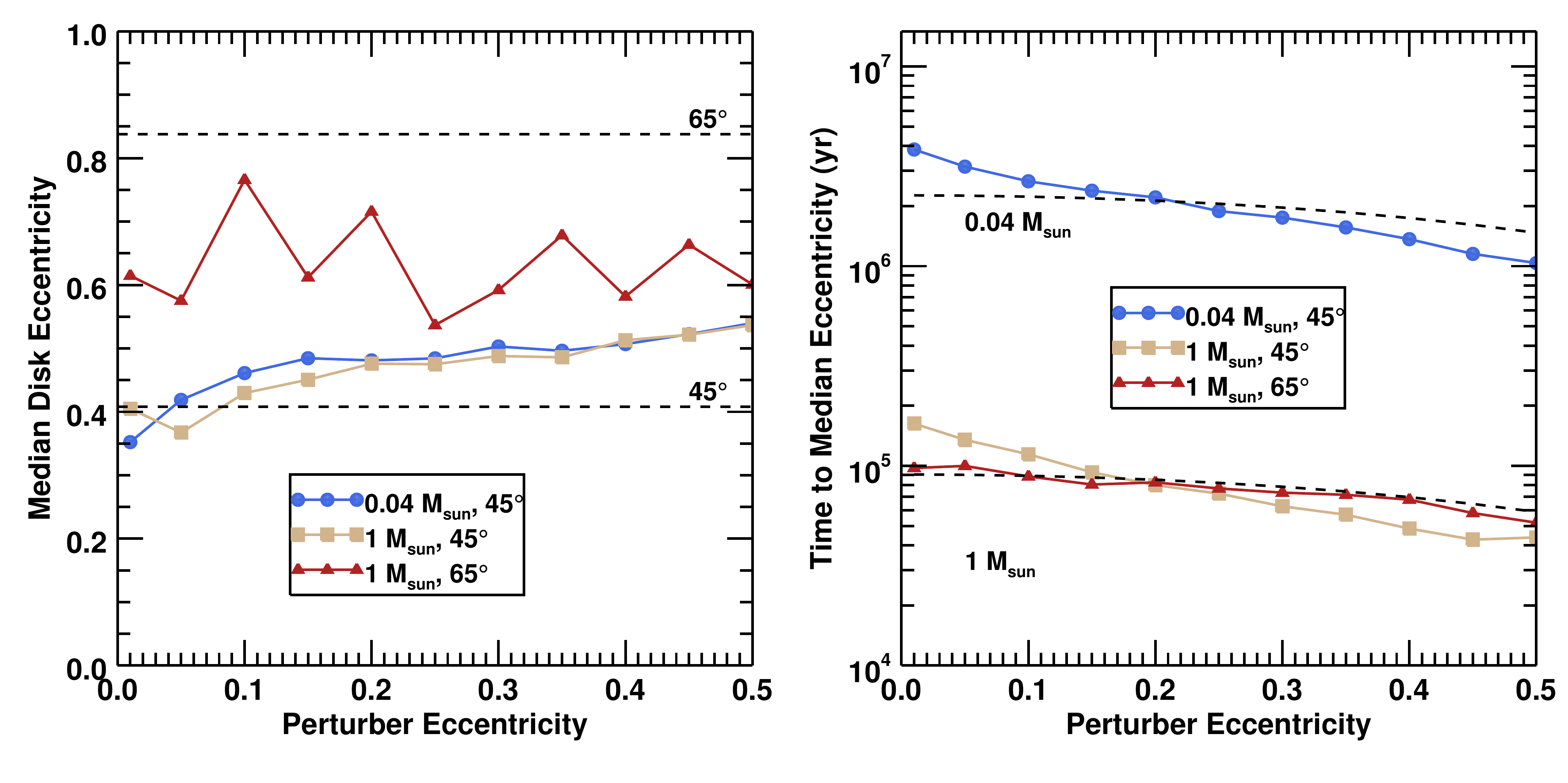}
	\caption{Left: Median eccentricity of the rings in the disk vs. the eccentricity of the perturber, for each of the three ring simulations. Dashed lines indicate maximum eccentricities predicted by Equation \ref{eq:emax}. Right: Time at which the average eccentricity of the rings reached the median eccentricity for each of the ring simulations. Dashed lines indicate the $t_{\rm Kozai}$ predicted by Equation \ref{eq:tKozai}.}
	\label{fig:masscomp}
\end{figure*}

As demonstrated by the overlap in the shaded regions in Figure \ref{fig:eccvstime}, the ring model, on average, nicely mimics the eccentricity output of the SMACK model, so we used the computationally faster ring code to test how well we can generalize our model. We ran a second simulation of the $220-250$ AU disk described in Table \ref{tab:initial}, changing the mass of the perturber to 1~M$_{\odot}$ to test the effect of perturber mass on the disk's eccentricity evolution. We also ran a third simulation of this disk, with a 1~M$_{\odot}$ perturber inclined to $65^{\circ}$ rather than $45^{\circ}$, to test the effect of perturber inclination. Our results are summarized in Figure \ref{fig:masscomp}.

We compared the results presented in Figure \ref{fig:masscomp} with the predicted values for maximum eccentricity $e_{max}$ and the time necessary to reach the maximum eccentricity, $t_{Kozai}/2$, using Equations \ref{eq:emax} and \ref{eq:tKozai}. These analytic predictions are shown as dashed lines in Figure \ref{fig:masscomp}. While the predicted timescales shown in the right panel of Figure \ref{fig:masscomp} agree well with our ring model results, the simulated disk eccentricities deviate substantially. This is an effect of higher-order terms of the Kozai-Lidov perturbations, such as the octupole terms that become significant in the case of an eccentric perturber \citep[see, e.g.,][]{Naoz2016} and the gravitational interactions between the rings. The Kozai-Lidov timescale predicted by Equation \ref{eq:tKozai} is derived from the quadrupole approximation, which neglects these terms. Since these higher-order terms become more significant for higher perturber and ring eccentricities, the simulation with the $65^{\circ}$ perturber deviates from the quadrupole prediction more significantly than the simulations with the $45^{\circ}$ perturbers.

Despite these differences between the simulation results and the analytic predictions, our ring models confirm that while the timescale of the Kozai-Lidov perturbations depends on the mass of the perturber, the median disk eccentricity under these perturbations does not. This indicates that our SMACK results concerning the effects of eccentricity excitation, presented in the following sections, could be generalized to larger perturbers, with only the timescale of the disk evolution varying with perturber mass.

\begin{figure*}[t]
	\centering
	\includegraphics[width=\linewidth]{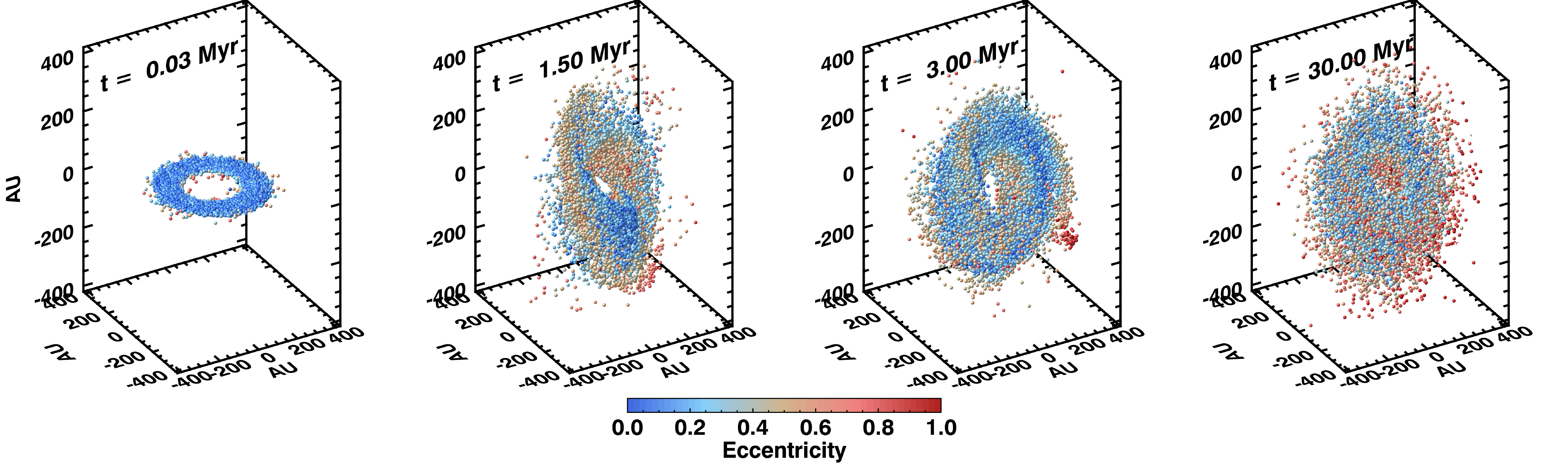}
	\caption{Time evolution of the 3D structure of the SMACK-simulated disk under the influence of an inclined exterior perturber. Colors indicate the eccentricity of the superparticles.}
	\label{fig:ecc3d}
\end{figure*}

Figure \ref{fig:ecc3d} shows the 3D spatial distribution of superparticles in the SMACK-simulated disk at various times in the simulation, with colors representing the eccentricity of each particle. As in the ring model (Figure \ref{fig:rings3d}), the Kozai-Lidov excitations from the perturber greatly increase the vertical extent of the disk, within 1.5 Myr for the SMACK simulation. The longitudes of ascending node of the disk particles also precess under Kozai-Lidov perturbations, creating a complex 3D structure. By 30 Myr, the most highly eccentric particles can be found in the outermost regions of the disk, closer to the perturber, while particles closer to the star have lower eccentricities. This is an effect of the presence of an exterior perturber, as particles closer to the perturber experience a shorter Kozai-Lidov timescale (Equation \ref{eq:tKozai}). We will compare this scenario to the case of an interior perturber in Sections \ref{sec:dust} and \ref{sec:sizedist}.

\subsection{Dust Production}
\label{sec:dust}

Fragmenting collisions between parent bodies in the debris disk will produce smaller dust grains, which are observable at infrared and visible wavelengths. During a simulation, SMACK tracks the redistribution of parent body sizes ($\ge 1$~mm) as the collisions grind them down to smaller sizes. The plots in Figure \ref{fig:ecc3d} show the 3D distribution of these larger parent bodies in the exterior brown dwarf disk. SMACK does not include radiative forces such as radiation pressure or Poynting-Robertson (PR) drag that affect smaller grains and therefore cannot track the orbits of bodies $<1$ mm in size, but it does simulate the collisions that produce these grains. The timescale for radiation pressure and PR drag to alter the orbit of a $\ge 1$~mm body is significantly longer than the collision timescale in these disks, and therefore we do not consider the effects of radiation forces on the parent bodies. These bodies are probed by millimeter and submillimeter observations; imaging at shorter wavelengths map the distribution of smaller grains, down to $\sim\mu$m sizes, below which grains are removed rapidly from the system by radiation pressure. Tracing the orbits of these $\mu$m-sized grains under the influence of radiative forces is beyond the scope of this work, but we can use SMACK to map where small grains are produced in the system.

\begin{figure*}[t]
	\centering
	\includegraphics[width=\linewidth]{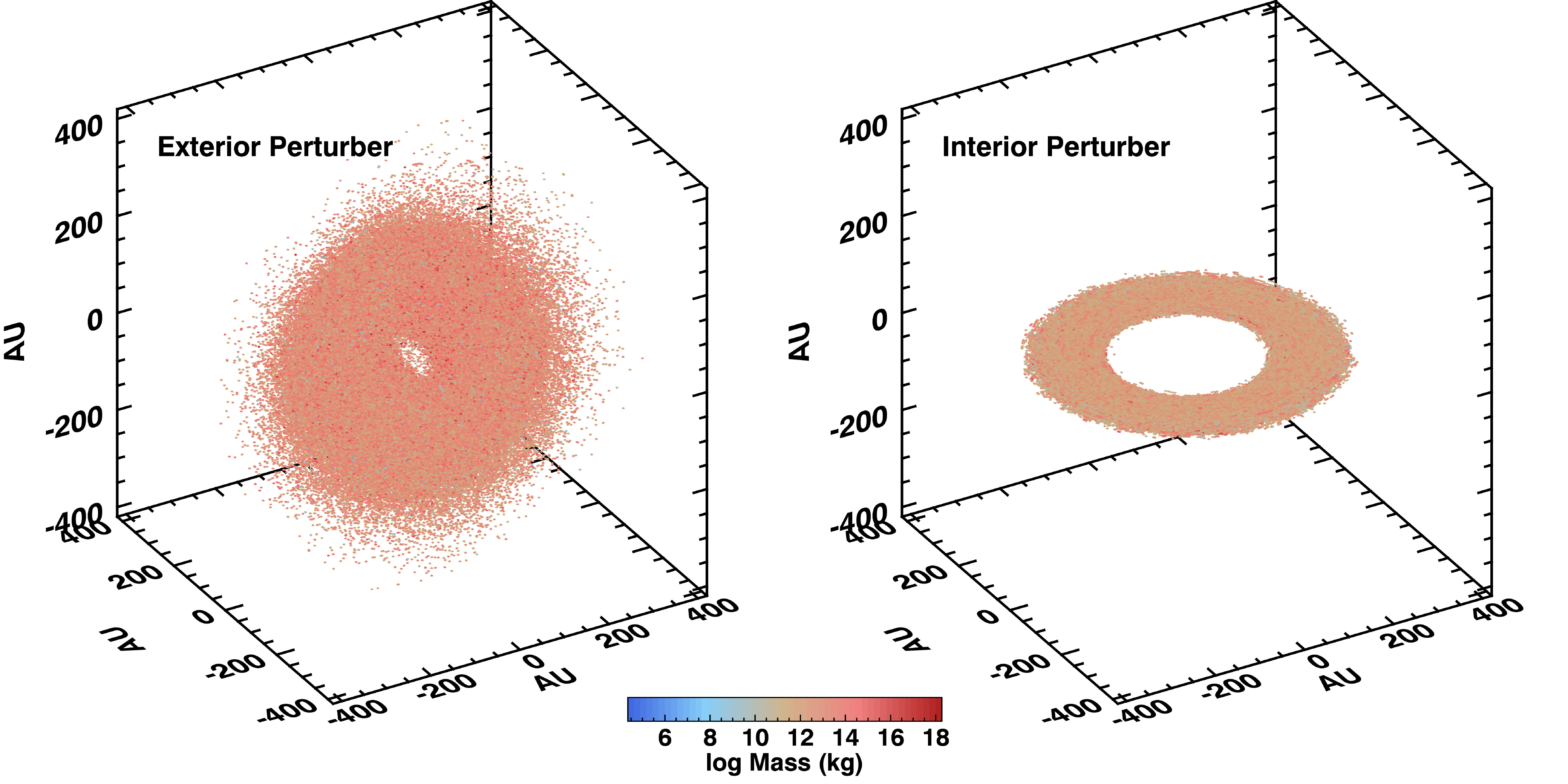}
	\caption{3D map of the dust produced in collisions in the SMACK disks with an exterior brown dwarf (left) and an interior planet (right) between 25 and 30 Myr. The spatial distribution of the dust in the exterior brown dwarf simulation reflects the shape of the parent body disk at 30 Myr, shown in Figure \ref{fig:ecc3d}.}
	\label{fig:dust3d}
\end{figure*}

We compared our nominal system with the disk configuration shaped by perturbations from an interior perturber. This scenario is often invoked in the literature to explain the presence of small dust particles \citep[e.g.,][]{Smith1984, Backman1993, Zuckerman2004, Kalas2005, Schneider2009}. In Figure \ref{fig:dust3d} we plot the locations of collisions that occurred between 25 and 30 Myr in each disk, scaled by the mass of dust produced in each collision. In the final 5 Myr of the simulations, the two disks produced similar amounts of dust: $2.1\times10^{-12}~\hbox{M}_{\odot}$ and $3.2\times10^{-12}~\hbox{M}_{\odot}$ in the exterior and interior perturber cases, respectively. While we would expect the orbits of the dust created in these collisions to evolve immediately under the influence of radiative forces, Figure \ref{fig:dust3d} provides an illustration of the gross differences between the shape of dust disks produced via an exterior Kozai-Lidov perturber versus an interior eccentric perturber. 

The spatial distribution of the dust produced in the disk with the exterior brown dwarf (left panel of Figure \ref{fig:dust3d}) reflects the shape of the excited parent body disk at 30 Myr, shown in the last panel of Figure \ref{fig:ecc3d}. Because a Kozai-Lidov perturber excites the inclinations of the parent bodies as well as their eccentricities, the parent body disk (and therefore the dust disk in Figure \ref{fig:dust3d}) is puffy and extremely vertically extended. By contrast, the disk with the interior planet (right panel of Figure \ref{fig:dust3d}) remains flat after 30 Myr, which is unsurprising given that we set the initial orbit of the planet to be coplanar with the disk. As a result, the parent bodies in this disk have little to no free inclinations, and remain in the plane of the planet's orbit. 

Is it possible for a planet orbiting interior to the disk to excite the inclinations of the parent bodies and produce a puffy disk similar to the disk with the exterior brown dwarf? The most well-studied planet orbiting with a non-zero inclination relative to its disk is $\beta$ Pictoris b, and indeed, the warp observed in the $\beta$ Pictoris disk \citep[e.g.,][]{Kalas1995, Heap2000, Golimowski2006} was attributed to the inclination of the planet \citep{Mouillet1997, Augereau2001, Dawson2011} before $\beta$ Pictoris b was even detected by \citet{Lagrange2010}. However, the inclination of $\beta$ Pictoris b relative to the disk is small enough ($<10^{\circ}$) that the maximum vertical extent of the disk is only $\sim10$ AU at a distance of $\sim100$ AU from the star \citep{Heap2000, Golimowski2006}. We attempted to reproduce a puffy disk with an inclined planet by running the same interior planet simulation described in Section \ref{sec:smack}, but setting the planet's inclination to $30^{\circ}$. However, after 30 Myr, only $10\%$ of the superparticles survived, indicating that a planet inclined enough to create a puffy disk will instead dynamically destroy the disk on a relatively short timescale, while a planet with a small enough inclination to allow the disk to survive (e.g., $\beta$ Pictoris) will not excite a large vertical extent in the disk.

Conversely, could an exterior Kozai-Lidov perturber produce a flatter disk, resembling a disk produced by an interior perturber? Exciting eccentricity via the eccentric Kozai-Lidov mechanism for low inclination or coplanar system would require an initially eccentric inner orbit \citep{Li2014}. Otherwise, a distant perturber with a small inclination $(\lesssim 40^{\circ})$ will not excite the eccentricities of the disk and trigger a collisional cascade, since the eccentric Kozai-Lidov mechanism produces large eccentricity oscillations only for systems with high initial inclinations \citep[e.g.,][]{Naoz2016}.  Moving the perturber closer, such that the hierarchical approximation breaks, tends to destroy the disk, as the excited eccentricity will quickly reach unity. 

Thus, we conclude that Figure \ref{fig:dust3d} illustrates one observable difference between a dust disk generated by an exterior perturber and one produced by an interior perturber: the disk with an exterior perturber will be extremely vertically extended. This spreading may make the dust fainter and more difficult to detect, but if an observed dust disk can be resolved, its vertical extent can indicate the type of perturber exciting the disk.
	
\subsection{Size Distribution Evolution}
\label{sec:sizedist}

\begin{figure*}[t]
	\centering
	\includegraphics[width=\linewidth]{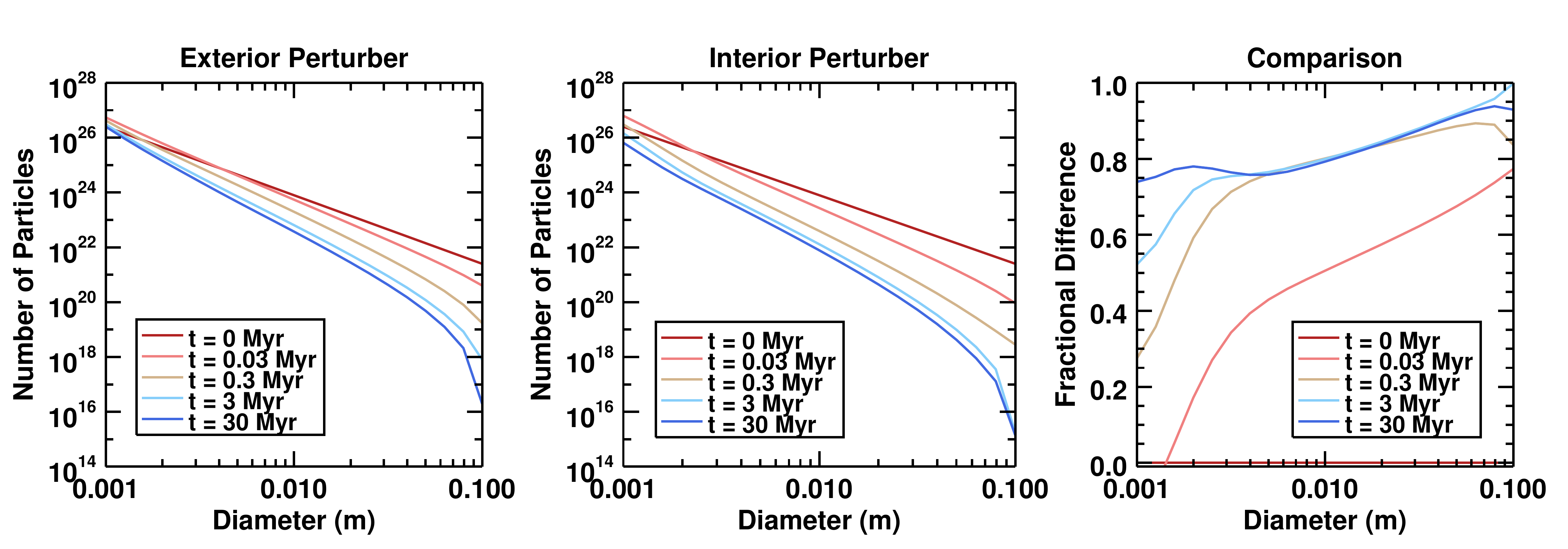}
	\caption{Total size distribution of the superparticles in the SMACK disk at various times in the exterior perturber (left) and interior perturber (center) simulations. As particles collide and fragment, the mass in larger bodies is redistributed into smaller bodies. The population of small grains is continually replenished while the population of larger grains is not, so the size distribution becomes steeper with time. We also plot the fractional difference between the two simulations over time (right), defined as $(n_{\rm ext}-n_{\rm int})/n_{\rm ext}$.} 
	\label{fig:sizedisttime}
\end{figure*}

As the parent bodies in the disk collide, the size distribution will evolve as larger bodies are broken and smaller fragments are produced. In Figure \ref{fig:sizedisttime} we plot the size distribution of both SMACK-simulated disks (with an exterior brown dwarf and with an interior planet) at various times during the simulations. We set the initial size distribution of the disk to be an incremental power law with index $-2.5$ (for logarithmic size bins). In an infinite collisional cascade with a constant velocity distribution, this index of the size distribution of the disk would remain constant \citep{Dohnanyi1969}. However, because we do not provide a source of replenishment for the largest bodies, the size distributions of both simulated disks become steeper as the collisions excited by the perturber process the mass in the parent bodies into smaller grains. 

There is no obvious difference in the shape or slope of the size distributions at in the two simulations, so in the rightmost panel of Figure \ref{fig:sizedisttime} we plot the fractional difference, $f$, between the size distributions of the two simulations at various times, where we define $f = (n_{\rm ext}-n_{\rm int})/n_{\rm ext}$, where $n_{\rm ext}$ and $n_{\rm int}$ are the size distributions of the simulations with the exterior and interior perturbers, respectively. Values of $f$ close to zero indicate near-agreement between the size distributions of the two simulated disks, while larger values indicate that $n_{\rm ext}>n_{\rm int}$. The total mass represented by each size distribution is a function of the collision rates in the disks and thus the specific configurations of the two systems; for example, the disk with the interior perturber loses roughly an order of magnitude more particles than the disk with the exterior perturber. While modeling the mass loss for an individual disk is challenging in the absence of information about the initial mass of the disk, future work simulating the time evolution of a suite of disks with interior or exterior perturbers may help interpret the measured size distributions of a large sample of disks.

Examining the shape of the $f$ curve at various times can provide more general insights into the different evolutions of these two scenarios. At earlier times in the simulations, the slopes of their size distributions differ, indicated by the steep slope of $f$, but this slope becomes shallower as the disks evolve. This indicates that the dynamics of the exterior perturber preserve larger bodies compared to an interior perturber, especially early in the evolution of the system. The deviation of the $f$ curves from a straight line also indicate a difference in the shape of the size distributions of the disks.

Despite these fine details in the fractional difference in the size distributions, our results indicate that both an exterior Kozai-Lidov perturber and an interior eccentric perturber will collisionally process the parent bodies in a disk and, in the absence of larger bodies to replenish the mass lost in the collisional cascade, change the slope of the size distribution. It would be challenging to distinguish between the exterior and interior perturber scenarios for a given disk by simply measuring the total size distribution of the disk, especially if the age of the system is not well-constrained. 

However, the timescale of this collisional processing should vary with location in the disk because the secular timescale depends on the distance to the perturber for both the eccentric planet scenario and the Kozai-Lidov scenario. In the case of an interior planet, the parent bodies on the inner edge of the disk will experience a shorter secular timescale than the parent bodies on the outer edge, due to their proximity to the planet. This is reversed for the exterior brown dwarf disk: the parent bodies on the outer edge of the disk will experience a shorter Kozai-Lidov timescale (Equation \ref{eq:tKozai}) than parent bodies orbiting closer to the star. This is illustrated in the last panel of Figure \ref{fig:ecc3d}, which shows that the most eccentric particles can be found at the outer edges of the Kozai-Lidov-excited disk. Variations in the velocity distribution of colliding particles within in a disk can alter the size distribution of the collisional cascade \citep{Pan2012}, so the difference in eccentricities at the inner and outer edges of the disk could result in different size distributions in these regions. 

\begin{figure}[t]
	\centering
	\includegraphics[width=\columnwidth]{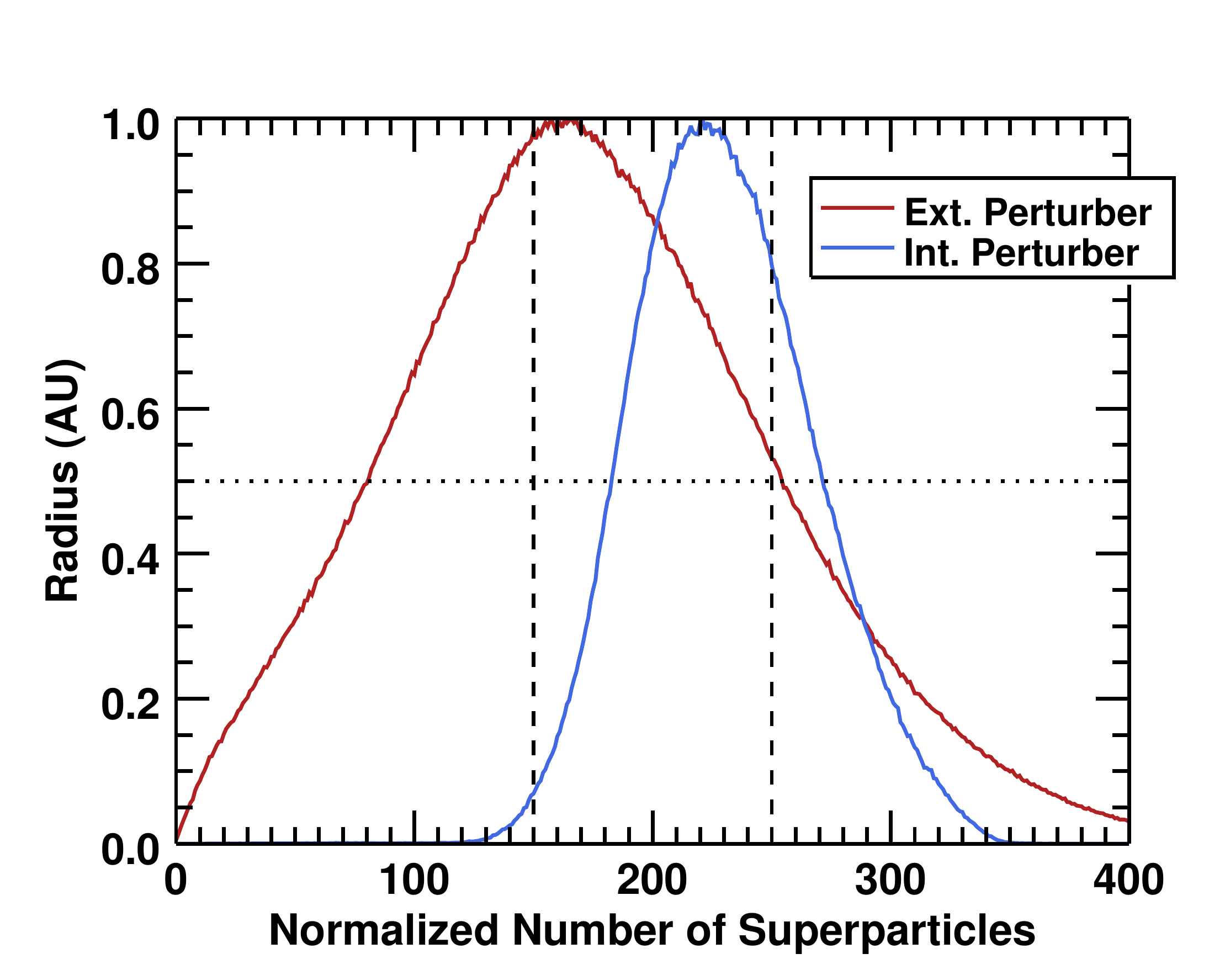}
	\caption{Radial distributions of superparticles in the exterior brown dwarf and interior planet simulations. The vertical dashed lines show the initial distribution of each ring. The horizontal dotted line illustrates the half-max point at which the edges of the disk were measured.}
	\label{fig:radialcut}
\end{figure}

To test this hypothesis, we measured the size distribution at the inner and outer edges of each simulated disk at 30 Myr. Figure \ref{fig:radialcut} shows the normalized histograms of the superparticle radii in each disk at 30 Myr. Note that both disks have shifted from their initial uniform distribution between $150-250$ AU, shown with vertical dashed lines. The disk with the exterior brown dwarf perturber broadened and the peak shifted inwards, while the peak of the disk with the interior planet shifted outwards slightly. To measure the edges of the disk, we used the half-maximum radius \citep[e.g.,][]{Chiang2009, Nesvold2015} and defined the inner and outer edges as the minimum and maximum radii, respectively, where the radial distribution of the superparticles reached half its maximum value. This half-max point is illustrated with a horizontal dotted line in Figure \ref{fig:radialcut}. Using this technique, we located the inner and outer edges of the exterior brown dwarf disk at $\sim77$ and $\sim253$ AU, respectively, and the inner and outer edges of the interior eccentric planet disk at $\sim182$ and $\sim272$ AU, respectively.

\begin{figure*}[t]
	\centering
	\includegraphics[width=\linewidth]{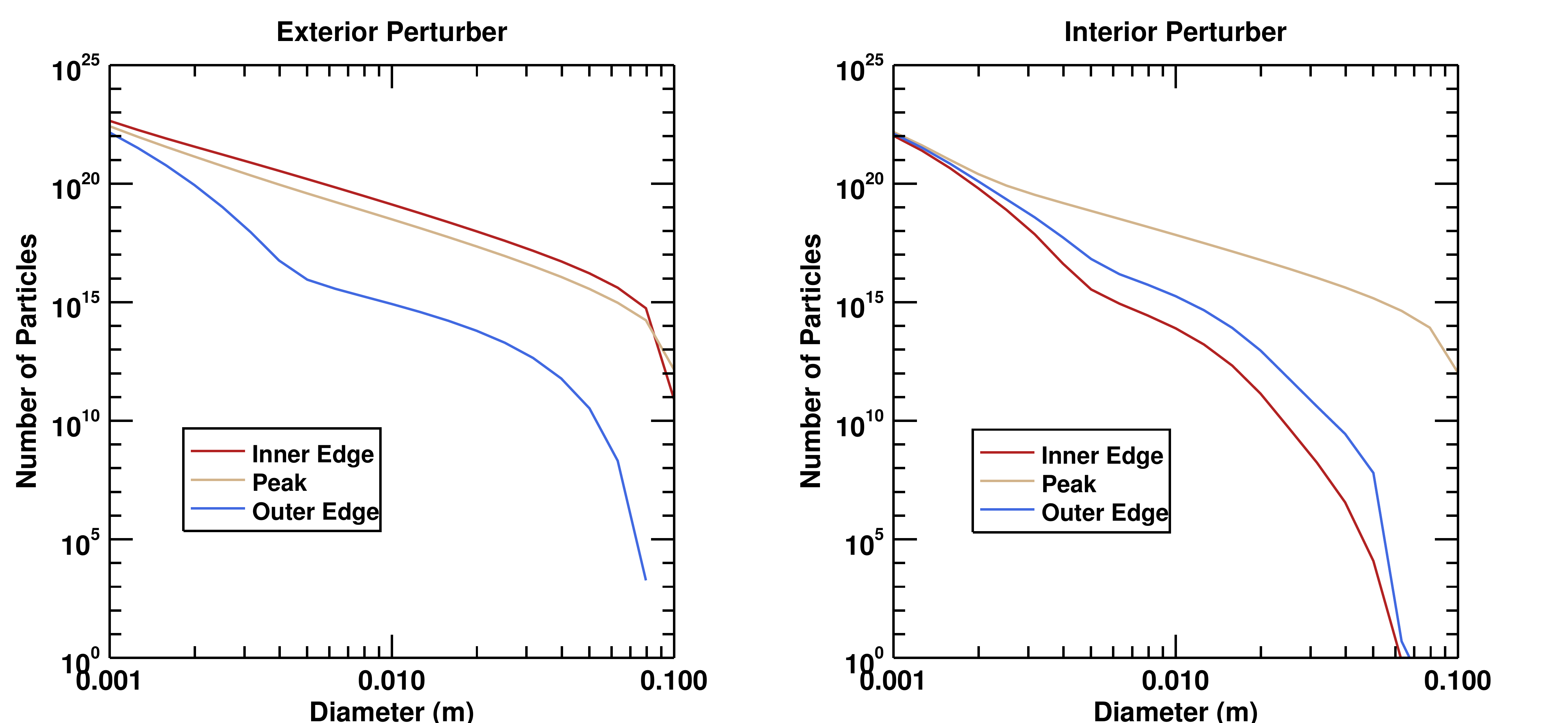}
	\caption{Average size distribution of superparticles in the SMACK disk at 30 Myr, for superparticles orbiting at the inner and outer edges of the disk with the exterior perturber (left) and the disk with the interior perturber (right).}
	\label{fig:sizedistradius}
\end{figure*}

To measure the size distribution at the inner and outer edges of the disks, we calculated the average size distributions of all superparticles located at a radius within 5 AU of each edge at 30 Myr for each disk. In Figure \ref{fig:sizedistradius} we plot these size distributions, as well as the size distributions of the $50\%$ of superparticles orbiting nearest the peak of each disk. In the disk with the exterior perturber (shown in the left panel of Figure \ref{fig:sizedistradius}), the inner edge of the disk and the central peak of the disk exhibit roughly the same average size distribution, roughly a power law with a sharp drop at the largest size bin. However, the size distribution at the inner edge is significantly different, a broken power law indicating a much higher level of mass loss and collisional processing. These results support the hypothesis that the parent bodies orbiting at the outer edge of the disk have experienced more collisions due to their shorter Kozai-Lidov timescale.

Conversely, in the disk with the interior perturber (shown in the right panel of Figure \ref{fig:sizedistradius}), only the central portion of the disk exhibits a power law size distributions. The parent bodies at both the inner and outer edges of the disk have broken power laws and evidence of significant collisional processing. While both edges of the disk appear to have experienced high collision rates, the inner edge exhibits a higher amount of mass loss by $\sim2-3$ orders of magnitude.

\section{Discussion}
\label{sec:discussion}

The detection of a dust disk can prompt the search for the massive perturber assumed to be exciting the disk. However, we have demonstrated that the body responsible for perturbing the disk could be orbiting interior to the disk or at a large distance exterior to the disk. Determining the nature of the perturber using observations of the disk would allow observers to constrain their search region. The results of our simulations indicate two methods for distinguishing between a resolved dust disk formed by an eccentric perturber orbiting interior to the disk and one formed by an inclined perturber orbiting exterior to the disk via the Kozai-Lidov mechanism. 

As discussed in Section \ref{sec:dust}, a Kozai-Lidov perturber will produce an extremely vertically extended disk by exciting the inclinations of the disk particles with a timescale that varies with radius (Figure \ref{fig:dust3d}). Imaging the disk at an orientation other than face-on will allow the observer to distinguish between an interior and an exterior perturber through this criterion alone. Another possible distinguishing criterion for resolved disks is that the sharp edges in a dust disk that are characteristic of a disk with an interior perturber \citep[e.g.,][]{Chiang2009, Lagrange2012a, Boley2012} will not be present in a disk perturbed by a distant, exterior companion.  

The 3D structure of the vertically extended Kozai-Lidov disk may result in observable spectral features even if the disk is viewed face-on. For example, \citet{Chiang1997} modeled the emission from circumstellar disks around T Tauri stars and derived emission features that indicate when the observed disk is flared. \citet{Jura2003} used the \citet{Chiang1997} expression for the effective temperature of a flat, optically thick disk to derive the predicted flux from the disk as a function of disk radius and the observed inclination of the ring. The maximum flux will occur in the case of a face-on ring, indicating that there exists a maximum value of $L_{\rm IR}/L_*$ that can be achieved for a flat disk with a given semi-major axis and radial width. A measured $L_{\rm IR}/L_*$ larger than this maximum value would indicate a vertically extended disk. The disks discussed in this paper are optically thin, and are therefore not directly comparable to the optically thick disks discussed by \citet{Chiang1997} and \citet{Jura2003}. However, radiative transfer simulations comparing the flat disk to the puffy disk in the Kozai-Lidov case, which are beyond the scope of this paper, may suggest additional features in the disk SEDs that could help distinguish between these scenarios.

If the disk appears face-on, Section \ref{sec:sizedist} suggests a second method for determining the nature of the perturber: comparing the size distributions at the inner and outer edges of the disk. Measuring the size distribution of an observed disk poses a challenge, but recent analyses of submillimeter disk observations \citep[e.g.,][]{Ricci2010, Ricci2012, MacGregor2015, Steele2015, Ricci2015a, Ricci2015b} demonstrate a promising technique for calculating the slope of the size distribution, $q$, using the dust opacity spectral index, $\beta$. At long ($\sim$mm) wavelengths, the dust opacity in a disk has a power-law dependence on frequency, $\kappa_{\rm dust} \propto \nu^{\beta}$, and the spectral energy distribution (SED) is $F_{\nu} \propto \nu^{\alpha_{\rm mm}}$, where the millimeter spectral index is $\alpha_{\rm mm} = \alpha_{\rm Pl} + \beta$, where $\alpha_{\rm Pl}$ is the index of the Planck function. \citet{Draine2006} derived the following relationship between $\beta$ and $q$: $\beta \approx (q-3)\beta_{\rm S} $, where $\beta_{\rm S}$ is the dust opacity spectral index in the small particle limit. This relationship is valid for size distributions with $3<q<4$. Thus, the index of the size distribution can be estimated as 
\begin{equation} \label{eqn:draine} q = \frac{\alpha_{\rm mm}-\alpha_{\rm Pl}}{\beta_{\rm S}}, \end{equation}
for $3<q<4$. For example, \citet{Ricci2012} analyzed sub-mm and mm observations of the Fomalhaut debris disk by estimating $\alpha_{\rm Pl} = 1.84\pm0.02$ and $\beta_{\rm S} = 1.8\pm0.2$, measuring the spectral index $\alpha_{\rm mm}$ of the SED, and calculating the grain size distribution to be $q=3.48\pm0.14$. 

Based on this technique, we propose the following method to determine the nature of the perturber of a resolved, face-on dust disk. First, perform multi-wavelength imaging of the disk to measure the SED at the inner and outer edges of the disk, then use the relationship between $q$ and $\beta$ in Equation \ref{eqn:draine} to estimate the size distribution indices $q_{\rm in}$ and $q_{\rm out}$. According to Figure \ref{fig:sizedistradius}, a measurement of $q_{\rm in} > q_{\rm out}$ indicates that the perturber is orbiting interior to the disk, while a measurement of $q_{\rm in} < q_{\rm out}$ indicates that the perturber is orbiting exterior to the disk, and perturbing the disk via the Kozai-Lidov mechanism. 

Note that both of these techniques require that the parent body disk be wide enough that the perturbation timescale is significantly different at the inner and outer edges of the disk. A narrow disk perturbed by an exterior perturber will not have a wide enough range of maximum inclinations to become significantly vertically extended. Also, the similar perturbation timescales in a narrow disk will result in similar collision rates, resulting in indistinguishable size distributions at the inner and outer edges of the disk.

Another intriguing difference between a population of second-generation dust produced via an interior, eccentric perturber and one produced via an exterior, Kozai-Lidov perturber, is that the parent body disk with the interior perturber will eventually evolve into an eccentric ring, apsidally aligned with the perturber's orbit \citep[e.g.,][]{Wyatt2005a}. At this point, the frequency of orbit crossing between parent body will decrease, resulting in a dynamically cold disk and greatly reducing the dust production rate. However, the oscillating parent body eccentricities induced by the Kozai-Lidov perturber will maintain the collision rate indefinitely, so the dust production in such a disk will be limited only by the mass available in the disk. Future simulations should investigate this effect on the lifetimes of the observable dust disks in these different types of systems.

\section{Summary and Conclusions}
\label{sec:summary}	

We modeled the perturbation of a circumstellar disk of particles by an exterior, inclined perturber. We demonstrated that the eccentricity evolution of the disk does not depend on the mass of the perturber except via the Kozai-Lidov timescale, so we focused our modeling efforts on a system with a 40~M$_{\rm Jup}$ perturber. The perturber excites the eccentricities and inclinations of the particles via the Kozai-Lidov mechanism, increasing the rate of collisions between particles and producing a dust population. We compared our results with a model of an eccentric, 10~M$_{\rm Jup}$ planet orbiting interior to the disk. While both the exterior and interior perturbers will excite dust-producing collisions between parent bodies, our results indicate two observable disk features that can be used to distinguish between an exterior and an interior perturber. First, the disk perturbed by an exterior companion will be significantly puffier and more vertically extended than a disk perturbed by an interior companion. We tested the possibility of an interior, inclined planet producing a vertically extended disk and found that such a configuration will destroy the disk. Second, comparing the index of the size distribution at the inner and outer edges of the disk can determine whether the perturber is orbiting exterior or interior to the disk. 

Besides providing a method to determine the nature of the perturber in a dust disk system, our models could also be used to analyze systems with known exterior perturbers. For example, \citet{Mawet2015} discovered a brown dwarf-mass object orbiting exterior to the warm disk around HR 3549, a 230 Myr-old A0V star. Although the disk is not resolved, further observations to constrain the orbit of the companion and detailed dynamical and SED modeling of the dust could reveal the connection, if any, between the production of short-lived dust and the perturbations from the companion. 

Future work should also explore the parameter space of an exterior perturber exciting a disk via the Kozai-Lidov mechanism. Models of protoplanetary disks have indicated that for disk masses greater than a few percent of the mass of the star, the Kozai-Lidov mechanism will be suppressed by self-gravity within the disk. We varied the disk mass in our ring simulations by an order of magnitude and found no qualitative difference in our results, but future simulations could search for the quantitative effects of disk self-gravity on the Kozai-Lidov mechanism for debris disk masses.

Note from Equation \ref{eq:tKozai} that the Kozai timescale also depends upon the mass of the perturber relative to the star and on the perturber's period and eccentricity. A suite of simulations varying the mass and orbit of the perturber could explore the effects of these parameters on the dust production rate and the structure of the dust disk. 

Enhanced collision rates, such as those produced via the Kozai-Lidov mechanism, can also alter the morphology of disks by eroding away the population of parent bodies \citep{Nesvold2015a, Nesvold2015}. In a future paper, will will address the effects of a Kozai-Lidov perturber on the lifetime of a disk due to collisional erosion, as well as the effects of enhanced collisions on planet formation in the disk.

\vspace{1em}

The authors wish to thank B. Zuckerman and H. Schlichting for useful discussions, and M. Fitzgerald and D. Mawet for helpful comments on the paper draft. SMACK simulations were run on the NASA Center for Climate Simulation's Discover cluster. Erika Nesvold was supported by the Carnegie DTM Postdoctoral Fellowship. Smadar Naoz would like to acknowledge partial support from the Sloan Foundation Research Fellowship as well as the Annie Jump Cannon Prize.

\bibliographystyle{../apj}
\bibliography{../Libraries/BrownDwarf}

\end{document}